# Noise Analysis of Current Sensitive Preamplifiers and Influence on Energy Resolution of NaI:Tl Detector System

Lin Jiang, Jinfu Zhu, Tao Xue, Jingjun Wen, Liangjun Wei, Jianmin Li and Yinong Liu

*Abstract*—Current Sensitive Preamplifiers (CSPs) are widely used in front-end electronics in data acquire system (DAQ), due to their ability to amplify signals directly. The optimization of energy resolution requires CSPs with suitable parameters such as gain, bandwidth, and low noise, etc. With the rapid improvement of CSP's bandwidth and trans-impedance gain, it is necessary to know how does the output noise of CSPs effect the energy resolution. For this purpose, we built the noise model of CSPs and analyzed the output noise voltage density of CSPs based on different commercial operation amplifiers (OP AMPs). Theoretical and experimental results shown that OP AMPs with low input noise voltage density is superior than others on energy resolution. Moreover, noise normalization factor has been defined to verify the relationship between energy resolution and noise of CSPs. The experimental results shown that the energy resolution and noise normalization factor are linearly conformable. This paper can provide reference on selecting commercial OP AMPs to design CSPs for optimization of energy resolution.

## I. INTRODUCTION

CURRENT sensitive preamplifiers (CSPs), also known as trans-impedance amplifier (TIA), are widely used for energy resolution measurement and PSD (Pulse Shape Discrimination). The CSPs convert the input current from PMT into voltage, so the output pulse can be digitized directly. [1] [2] [3] [8]. Various parameters of CSPs such as bandwidth, gain, stability and noise etc. should be taken into consideration for the optimization of energy resolution and PSD performance. Therefore, we should have some knowledge about the selection of OP AMPs and the parameters of the surrounding circuit when we design CSPs. The noise model of CSPs has been built to compare the output voltage density with the simulation and measurement. Moreover, the impact of noise on energy resolution was focused. The linear relationship between the energy resolution and the noise normalization factor have been derived and verified by the actual energy resolution of $^{40}$K based on different OP AMPs.

This paper will design and compare the actual performance of different OP AMPs. Section II describes the noise model of CSPs and theoretical calculation of energy resolution for NaI:Tl detector. Section III will introduce experiments set-up and results by means of different OP AMPs. The conclusion and future work will be discussed in Section IV.

It can provide reference for OP AMP chips selections based on CSPs performance and physical requirements.

## II. METHOD

### A. Noise Model of CSPs

The CSPs are similar to charge sensitive preamplifiers (CSAs) in circuit topology. The circuit model [5] for the electronics noise of operational amplifier is shown as in Fig. 1.

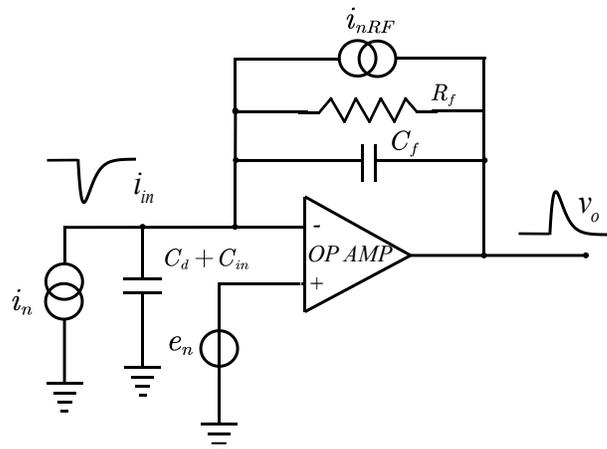

Fig. 1. OP AMP based current sensitive preamplifier with relevant noise generators.

$i_n$ and $e_n$ are the input current noise and voltage noise of amplifier, respectively. $i_{nRF}$ is Johnson noise of feedback resistor $R_f$. $C_d$, $C_{in}$ is detector junction capacitance and input capacitance of amplifier, respectively. As we all know, $i_n$ and $e_n$ of amplifiers have increasing noise at low frequencies, which called 1/f noise. The 1/f noise can be neglected because the CSP's bandwidth is more than 10 times the corner frequencies. Therefore, the input noise sources are equivalent to white noise generators [2] [4].

At the input terminal, current noise sources are connected in parallel, while voltage noise sources are connected in series. An input current noise produces an output voltage noise, so the gain is simply $-R_f$. Different noise sources can be equivalent to the input terminal, represented by the total input noise current density. For the input current noise $i_n$ of amplifier and the Johnson noise on feedback resistor, it contributes directly to

Manuscript received October 22nd, 2020. This work was supported by the National Natural Science Foundation of China (U1865205).

The authors are with the Department of Engineering Physics, Tsinghua University, and also with the Key Laboratory of Particle& Radiation Imaging (Tsinghua University), Ministry of Education, Beijing, 100084, China (Corresponding author is Tao Xue, e-mail: xuetao@mail.tsinghua.edu.cn).

the input noise current density. However, for the input voltage noise $e_n$ of amplifier, it generates current noise through the input capacitor and feedback resistor, as "$e_nC$" noise [6], which can be defined as $i_{nc}$ and $i_{nf}$ as shown in equation (1) and (2), respectively.

$$i_{nc} = e_n\omega C_{in} = 2\pi e_n C_{in} f \tag{1}$$

$$i_{nf} = e_n/R_f \tag{2}$$

Johnson noise of feedback resistor is flat with frequency, with a current noise $i_{nRF}$, as shown in equation (3).

$$i_{nRF} = \sqrt{4kT/R_f} \tag{3}$$

Where $k$ is Boltzmann's constant, $T$ is the absolute temperature in Kelvins (K=℃+273.16). So for a $1k\ \Omega$ feedback resistor at 25℃, $i_{nRF} = 4pA/\sqrt{Hz}$.

This four noise sources in the mathematically calculation are uncorrelated [1]. The proper additive formula for uncorrected noise is a root-sum-square (RSS). The total input noise current density can be calculated as equation (4). The noise current density at input terminal is shown in Fig. 2. The crossover frequency $f_X$ and the circuit's bandwidth $f_c$ is marked in fig. 2. and shown as equation (5) and (6).

$$i_{nt} = \sqrt{i_n^2 + i_{nRF}^2 + i_{nc}^2 + i_{nf}^2} \tag{4}$$

$$f_X = \frac{\sqrt{4kT/R_f}}{2\pi e_n C_{in}} \tag{5}$$

$$f_c = \frac{1}{2\pi R_f C_f} \tag{6}$$

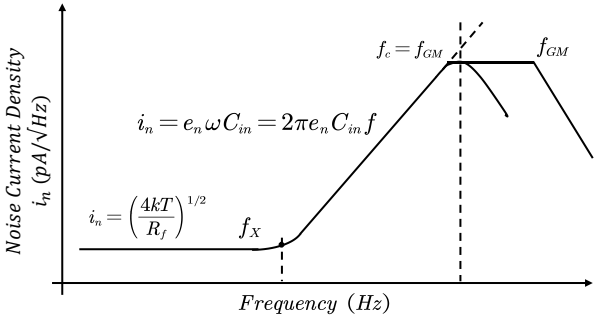

Fig. 2. Input-referred noise current density spectrum for the CSPs

### B. Theoretical Calculation of Energy Resolution

The energy resolution is contributed by two parts: the inherent energy resolution of the detector and the electronic noise, which is calculated by equation (7) from reference [7],

$$res_m^2 = res_{intrinsic}^2 + res_{noise}^2 = res_{intrinsic}^2 + \left(\frac{2.355\sigma_{Q_n}}{Q_t}\right)^2 \tag{7}$$

where $Q_t$ is central value of the integral charge, $\sigma_{Q_n}$ is the uncertainty of the integral charge.

The uncertainty of integral charge of real system is,

$$\sigma_{Q_n}^2 = \frac{1}{2} S_{x0} \cdot \left(1 + \frac{t_T}{t_s}\right) \cdot t_T \tag{8}$$

Where $t_T$ is the length of long gate, $t_s$ is the sample interval. The total noise power spectral density of the system $S_{x0}$ can be calculated in equation (9),

$$S_{x0} = \frac{\sigma_{total}^2}{A^2 \cdot F_s/2} \tag{9}$$

Where A is the trans-impedance gain of CSPs (k$\Omega$), $F_s$ is the sampling rate of ADC.

By substituting Eq. (8), Eq. (9) to Eq. (7), the expression for energy resolution can be simplified as Eq. (10),

$$res^2 = res_{intrinsic}^2 + K \cdot \frac{1}{Q_t^2} \cdot \frac{\sigma_{total}^2}{A^2} \tag{10}$$

In summary, there can be a linear model between the energy resolution and noise normalizing factor, $\frac{1}{Q_t^2} \cdot \frac{\sigma_{total}^2}{A^2}$.

## III. EXPERIMENT AND RESULTS

### A. Output Noise Voltage Density of CSPs

The output noise voltage density by means of simulation, computation and measurement in experiment are compared in Fig. 3. based on LTC6268, for example. It's important to note that the feedback resistance and capacitance is 1k Ohm and 1.8 pF, respectively.

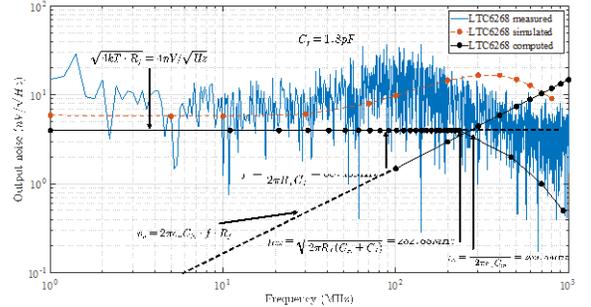

Fig. 3. Output-referred noise voltage density spectrum for the CSPs based on LTC6268

The noise current created by the amplifier's input voltage noise rises proportional to input capacitance, input voltage noise of OP AMPs and frequency. Fig. 2. shows the "$e_nC$" noise becoming dominant over other noise contribution. Considering the detector output capacitance (about 19.3pF), the input capacitance of OP AMPs can be ignored. Consequently, we can conclude that the OP AMP with less input voltage noise will have better performance on energy resolution. The experiment result shows that the energy resolution of $^{40}$K is better when the input voltage noise of amplifier is smaller. It's important to note that the waveform data are processed by low-pass filter (LPF), which -3dB bandwidth is 60MHz.

### B. Experiment Set-up and Energy Resolution Measurement

The detection system consists of a 127 mm diameter and 127 mm height NaI (Tl) scintillator, a B133D01 PMT from ADIT, CSPs based on different commercial OP AMPs and WavePro HD high definition oscilloscope from TELEDYNE LECROY.

To compare the performance between different OP AMPs and verify the linear model between energy resolution and noise normalizing factor, we select seven commercial OP AMPs (OPA657, OPA846, OPA847, OPA855, OPA858, LTC6268 and LTC6268-10) from Texas Instruments® and Analog Devices®, which are used for noise and energy spectrum measurement experiments based on NaI:Tl detector. Table I shows the specification of selected OP AMPs. Fig. 4. shows the energy spectrum with LTC6268, the energy resolution of $^{40}$K is 6.122% ± 0.204%.

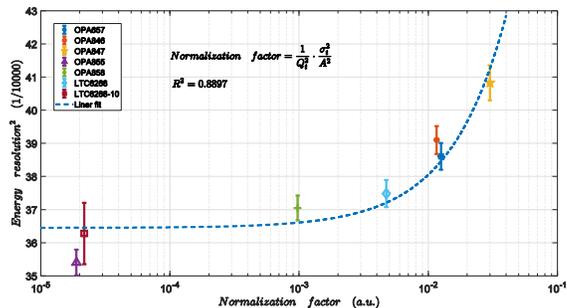

Fig. 5. Energy resolution measured by different OP AMPs with goodness of linear fit $R^2$=0.89

IV. SUMMARY AND FUTURE WORK

As we remarked earlier, the input voltage noise of OP AMPs is a critical parameter on the condition that the detector output capacitance is about almost twenty. Consequently, the High-frequency noise is dominated by "$e_n C$" noise, so, the bipolar input amplifiers (i.e. input bias current is two or three orders of magnitude large than MOSFET or JFET input, input voltage noise is small relatively) are more suitable than others (i.e. FET input). Moreover, amplifiers with low input capacitance are more suitable in sensors with low output capacitance (e.g. several pF).

This paper theoretically deduces the relationship between energy resolution and normalization factor, which is verified by 7 OP AMPs based on the NaI:Tl detector experiments.

In the future, more high performance commercial OP AMPs will be used for verify the model. More suitable OP AMPs will be selected in CSPs for different types of detectors such as CLYC, LaBr$_3$:Ce and NaI:Li. Besides, the parasitic capacitance of PCB (Printed Circuit Board) can be estimated by the noise spectral density to provide a reference for different plates.

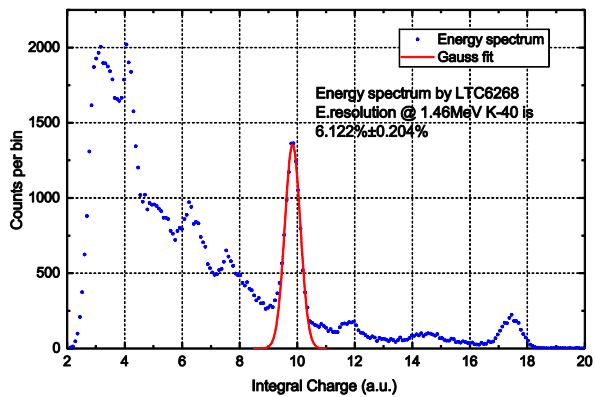

Fig. 4. Energy spectrum and gauss fit of $^{40}$K peak measured by LTC6268

TABLE I
SPECIFICATION OF SELECTED OP AMP

| OP AMP | $f_{GBP}$ (MHz) | $C_{in}$ / pF | $e_n$ ($nV/\sqrt{Hz}$) | $i_n$ ($fA/\sqrt{Hz}$) |
|---|---|---|---|---|
| OPA657 | 1600 | 5.2 | 4.8 | 1.3 |
| OPA846 | 1750 | 3.8 | 1.2 | 2800 |
| OPA847 | 3900 | 3.7 | 0.85 | 2500 |
| OPA855 | 8000 | 0.8 | 0.98 | 2500 |
| OPA858 | 5500 | 0.8 | 2.5 | -- |
| LTC6268 | 800 | 0.55 | 4.3 | 5.7 |
| LTC6268-10 | 4000 | 0.55 | 4.0 | 7 |

TABLE II
RESULT OF EXPERIMENT BASED ON DIFFERENT OP AMPS

| OP AMP | Noise std (mV) | Energy resolution (%) | $Q_t$ | A (k Ohm) |
|---|---|---|---|---|
| OPA657 | 0.07 | 6.2134 | 9.8193 | 1 |
| OPA846 | 0.17 | 6.2531 | 9.8335 | 1 |
| OPA847 | 0.17 | 6.3897 | 9.8759 | 1 |
| OPA855 | 0.25 | 5.9500 | 48.7816 | 10 |
| OPA858 | 0.46 | 6.0867 | 25.3639 | 5.11 |
| LTC6268 | 0.03 | 6.1224 | 9.8427 | 1 |
| LTC6268-10 | 0.14 | 6.0234 | 50.1969 | 5.1 |

The $^{40}$K energy resolution are calculated using different OP AMPs. The variance of the actual operational amplifier output noise is also calculated, which is used to calculate the noise normalizing factor. The experimental results are shown in Table II. The calculated normalization factor and energy resolution of different CSPs are shown in Fig. 5. The goodness of linear fit $R^2$ is ~0.89. It indicates that the energy resolution and noise normalization factor are linearly conformable.